\input harvmac.tex
\input epsf

\overfullrule=0pt
\parskip=3pt
 at 14truept
\overfullrule 0pt





\input mssymb
\input amssym.def
\input amssym.tex
\def\Z{\Bbb Z}\def\R{\Bbb R}\def\N{\Bbb N}



\def\hexnumber@#1{\ifcase#1 0\or1\or2\or3\or4\or5\or6\or7\or8\or9\or
        A\or B\or C\or D\or E\or F\fi }

%
\def\boxit#1{\leavevmode\kern5pt\hbox{
	\vrule width.2pt\vtop{\vbox{\hrule height.2pt\kern5pt
        \hbox{\kern5pt{#1}\kern5pt}}
      \kern5pt\hrule height.2pt}\vrule width.2pt}\kern5pt}


\def\eqalignD#1{
\vcenter{\openup1\jot\halign{
\hfil$\displaystyle{##}$~&
$\displaystyle{##}$\hfil~&
$\displaystyle{##}$\hfil\cr
#1}}
}
\def\eqalignT#1{
\vcenter{\openup1\jot\halign{
\hfil$\displaystyle{##}$~&
$\displaystyle{##}$\hfil~&
$\displaystyle{##}$\hfil~&
$\displaystyle{##}$\hfil\cr
#1}}
}


\def\\{\hfil\break}
\def\la{\lambda}

\def\y{{\tilde y}}
\def\xt{{\tilde x}}
\def\W{{\tilde W}}

\def\La{{\Lambda}}

\def\al{\alpha}
\def\be{\beta}
\def\ga{\gamma}

\def\+{\oplus}

\def\hfb{\hfill\break}
\def\Y{{\tilde Y}}
\def\y{{\tilde y}}

\font\huge=cmr10 scaled \magstep2
\font\small=cmr8


{\nopagenumbers
\rightline{December, 1998}
\bigskip
\centerline{{\huge\bf On Principal Admissible Representations}} 
\centerline{{\huge\bf and}} 
\centerline{{\huge\bf Conformal Field Theory{\footnote{$^\dagger$}{\small
Supported in part by NSERC.}}}} 
\bigskip\centerline{P. Mathieu\ and\ M.A.
Walton}
\bigskip

\centerline{{\it D\'epartement de Physique,
Universit\'e Laval}} \centerline{{\it Qu\'ebec (Qu\'ebec)\ \ 
Canada\ \ G1K 7P4}}
\centerline{pmathieu@phy.ulaval.ca}\bigskip
\centerline{{\it Physics Department,
University of Lethbridge}} \centerline{{\it Lethbridge, Alberta,
Canada\ \ T1K 3M4}}
\centerline{walton@uleth.ca}\bigskip

\leftskip=1cm \rightskip=1cm
\noindent{{\bf Abstract:}} The principal admissible representations of affine
Kac-Moody algebras are studied, with a view to their use in conformal field
theory. We discuss the generation of the set of principal admissible highest
weights, concentrating mainly on $A_r^{(1)}$ at rational level $k$. A  related 
algorithm is described that produces the Malikov-Feigen-Fuchs null vectors
of these representations. With the principal admissible description of the
highest weights, we are able to prove that field identifications (including
maverick ones) lead to the canonical description of the primary fields of the
nonunitary diagonal coset theories.  

\leftskip=0cm \rightskip=0cm

\vfill

\eject}

\pageno=1
\newsec{\bf Introduction}

The role in conformal field theory of integrable representations of affine
Kac-Moody algebras is well understood. These occur at positive integer level
$k$, and describe the excitations of primary fields in the Wess-Zumino-Witten
(WZW) conformal field theories. Integer level is required for a
single-valued Wess-Zumino contribution to the exponential of the WZW action. 

The integrable representations also figure in other conformal field 
theories, by the Goddard-Kent-Olive (GKO) coset construction. For example, the
unitary $W_n$ minimal models are described by the coset $A_{n-1,1}\oplus
A_{n-1,k}/A_{n-1,1+k}$, with $k\in {\N}$ (here $A_{n-1,k}$ denotes the affine
untwisted algebra $A_{n-1}^{(1)}$ at fixed level $k$). 

But there also exist nonunitary minimal models. In order to describe them by
the same GKO coset, we need to consider representations of affine algebras at
rational but noninteger (or fractional) level $k$. These 
admissible representations were discovered by Kac and Wakimoto 
\ref\kwone{V. Kac and M. Wakimoto, {\sl Proc. Nat. Acad. Sci. USA} {\bf 85}
(1988) 4956}\ref\kwtwo{V. Kac and M. Wakimoto, {\sl Adv. Ser. Math. Phys.} {\bf
7} (World Scientific, 1988) 138}, but their role in conformal field theory is
still being written. 

The admissible representations are nonunitary, except for the subclass of 
integrable representations. They nevertheless share many important properties
with the integrable representations. For
example, their characters obey a generalisation of
the Weyl-Kac character formula, and they have nice modular properties.  
This last feature is inherited by the nonunitary coset
conformal theories, as is necessary since they are  rational conformal 
field
theories, just like  their unitary cousins. 

In  \ref\mw{P. Mathieu and M.A. Walton, {\sl Prog. Theor. Phys.} (supplement)
No. 102 (1990) 229} we studied the modular properties of admissible
characters, and their consequences in the Verlinde formula and for
modular-invariant partition functions. In \ref\msw{P. Mathieu, D. S\'en\'echal
and M.A. Walton, {\sl Int. J. Mod. Phys.} {\bf A7}, supplement 1B (1992) 3255},
the use of admissible representations in diagonal GKO cosets was studied. For
example, a new class of intrinsically nonunitary field identifications
was discovered that showed, at least in many examples, that the canonical
description of the primary fields could be derived from the coset description.
One of our results here is the proof that this works generally. 

Key to the proof is the use of the so-called principal
admissible representations \ref\kwthree{V. Kac and M. Wakimoto, {\sl Acta Appl.
Math.} {\bf 21} (1990) 3}. Our main theme in this paper is that the description
of admissible highest weights as principal admissible weights is very natural,
and so should prove instrumental in clarifying the use of admissible
representations in conformal field theory. 

The main outstanding question does not have to do with GKO cosets, but rather
with the affine Kac-Moody algebras at fractional level themselves.
Certainly, naive use of the WZW action does not make sense at noninteger level
$k$.\foot{A partial geometric interpretation was given in \ref\fema{B.
Feigen and F. Malikov, {\sl Lett. Math. Phys.} (1994) 315}, however. But its
physical relevance is not clear.} But an action, and the geometrical
setting of the WZW action, are luxuries not accorded most rational conformal
field theories. One may still ask, therefore, if the affine algebras at
fractional level might nevertheless be the chiral algebras of sensible (albeit
nonunitary) rational conformal field theories. 

For the simplest case of $A_{1,k}$, some correlation functions have been
constructed  that obey the affine Ward-Takahashi identities and the
duality properties appropriate to a rational conformal field theory
\ref\pry{J.L. Petersen, J. Rasmussen and M. Yu, {\sl Nucl. Phys.} {\bf B481}
(1996) 577; P. Furlan, A.Ch. Ganchev and V.B. Petkova, {\sl Nucl. Phys}.
{\bf B491} (1997) 635}. Some progress has also been made in generalising these
results
\ref\pryr{J.L. Petersen, J. Rasmussen and M. Yu, {\sl Nucl. Phys.} {\bf B502}
(1997) 649;\hfb J. Rasmussen, {\sl Mod. Phys. Lett.} {\bf A13} (1998) 1281; 
{\rm hep-th/9807153}}. 

On the other hand, the fusion rules consistent with these $A_{1,k}$ results 
have not yet been reconciled with the Verlinde formula, which should hold for
all rational conformal field theories.

It is clear that if the Verlinde formula is to apply, some modification must be
made to the characters found by Kac and Wakimoto and so to their modular
matrices. The Kac-Wakimoto modular $S$ matrix in the Verlinde formula
gives some fusion coefficients equal to negative integers \ref\ks{I.G. Koh and
P. Sorba, {\sl Phys. Lett.} {\bf B215} (1988) 723}\mw. The pathology 
of the modular matrices also shows in other ways: modular invariants can be
constructed that would be the partition functions of theories with no identity
primary field \ref\lu{S. Lu, {\sl Phys. Lett.} {\bf B218} (1989) 46}. 

One suggestion \ref\bef{D. Bernard and G. Felder,
{\sl Comm. Math. Phys.} {\bf 127} (1990) 145} \ref\ay{H. Awata and Y. Yamada, {\sl Mod.
Phys. Lett.} {\bf A7} (1992) 1185} is that the negative Verlinde
coefficients are a sign (pun intended) that some of the highest-weight
representations should be reinterpreted as lowest-weight representations.
For the case $A_{1,k}$, this idea has motivated certain modifications of the 
affine characters and the modular $S$ matrices \ref\ram{S. Ramgoolam, {\it
New modular Hopf algebras related to rational k ${\widehat sl}(2)$},
hep-th/9301121.}\ref\imb{C. Imbimbo, {\it New modular representations and
fusion algebras from quantized $SL(2,\R)$ Chern-Simons theories},
hep-th/9301031} in order to produce non-negative integer Verlinde numbers. But
these prescriptions are not compelling to us; in particular,
they are not uniform as functions of the level denominator, 
and their generalisations to other algebras are not
clear. 

In \ref\fgp{P. Furlan, A.Ch. Ganchev and V.B. Petkova, {\sl Nucl. Phys}. {\bf
B518} (1998) 645.}\ref\fgptwo{P. Furlan, A.C. Ganchev and V.B. Petkova, 
{\it An extension
of the character ring of $s\ell(3)$ and its quantization}, math.QA/9807106}\ 
it is supposed that the
fusion eigenvalue matrix is nonsymmetric  in the
fractional-level case. The fusion algebra is conjectured for $A_{2,3/u-3}$, with
$u\in\N$ coprime to 3, by considering  $A_{2,k}$
null-vector-decoupling equations. It seems to have the sensible non-negativity
property. But the algebra is not reconciled with the Verlinde formula.    
 
Another possibility is that the theories with fractional-level affine
algebras as their chiral algebras (fractional-level theories, for short) are
not rational conformal field theories. Perhaps only a part of the full (chiral)
Hilbert space is represented by admissible representations. Other types of
representations would be transparent to the null-vector decoupling
calculations mentioned above \ay\fgp, and so may exist in the
fractional-level theories, so that they are non-rational \ref\mrg{M.R.
Gaberdiel, private communication}. 

Of course, one possibility is that the fractional-level theories simply do not
exist. We feel that the results of \pry\ are a strong indication to the
contrary, however.  

It is our hope that the generalisation of the $A_{1,k}$ results to other affine
algebras will help to clear up these issues, and this work is a 
contribution in that direction. Again, we believe that the principal
admissible description of highest weights will be important in this area. And
that the principal admissible description of highest weights is natural only
becomes clear in the more general context. 

In the next section, integrable, admissible and principal admissible
representations and their highest weights are reviewed. We also specify in
detail how to generate the full sets of principal admissible highest weights
for the $A_{r,k}$ algebras, at any rank $r$ and admissible level $k$. (We 
defer to an appendix the description of the $C_2$ and $G_2$
admissible highest weights.) In
section 3, the null vectors of the principal admissible representations are
described, in a way that connects directly with the results of section 2.
Section 4 deals with the nonunitary diagonal coset theories, and contains the
proof that the coset model leads to the canonical description of the primary
fields in the corresponding minimal models. Section 5 is a
conclusion.

\newsec{\bf Principal Admissible Representations}

\subsec{Affine untwisted algebras}

We first review the theory of integrable representations of affine
untwisted  Kac-Moody algebras. For the most part, we will 
use notation
already introduced by Kac and Wakimoto. 

Let $X_{r,k}$ denote the untwisted affine algebra $X^{(1)}_r$, at fixed level
$k$. For example, with $X=A$, we have the affine algebra sometimes indicated by
$\widehat{su}(r+1)_k$. Let $\Pi=\{\al_0,\al_1,\ldots,\al_r\, \}$ and $\Pi^\vee
=\{\al^\vee_0,\al^\vee_1,\ldots,\al^\vee_r\, \}$ be the sets of simple roots and
coroots of $X_{r,k}$, respectively. The corresponding set of fundamental weights
are $(\Pi^\vee)^*=:\Psi=\{\Lambda_j\ :\ j=0,1,\ldots,r\, \}$, so that
$(\al^\vee_j|\Lambda_\ell)=\delta_{j,\ell}$. The affine Weyl vector is defined
to be $\rho:=\sum_{i=0}^r\Lambda_i$. 

The simple roots $\alpha_j$ for $j=1,\ldots,r$ can be identified with the
simple roots of the horizontal subalgebra $X_r\subset X_{r,k}$. Generally, 
overbars will 
indicate objects associated with the finite-dimensional
horizontal subalgebra. So, $\bar\Pi=\{\al_1,\ldots,\al_r\, \}$ and $\bar\Pi^\vee
=\{\al^\vee_1,\ldots,\al^\vee_r\, \}$ are the sets of simple roots and coroots
of $X_r$. The `extra' affine simple root is $\alpha_0=\delta-\theta$, where
$\theta$ is the highest root of $X_r$, and $\delta$ is the basic imaginary
(null) root, i.e. $\{n\delta\ :\ 0\not=n\in \Z\, \}$ is the complete set of
imaginary roots. So 
\eqn\scroots{\Pi^\vee\ 
=\ \{\delta-\theta,\al^\vee_1,\ldots,\al^\vee_r\, \}\ \ .}
We have $\delta=\sum_{j=0}^r\,a_j\al_j$, where the $a_j$ are
the marks. We also use $\cal P$ as the operator that projects an
affine weight to the horizontal weight space, so that
${\cal P}\alpha_0=-\theta$ for instance.

Let
$\Delta,\Delta_+,\Delta^{re},\Delta^{\vee\,re},$
$\Delta_+^{re},\Delta^{\vee\,re}_+$ denote the sets of roots, positive
roots, real roots,  real coroots, positive real roots, and positive real
coroots, respectively. Then
$\bar\Delta,\bar\Delta^\vee,\bar\Delta_+,\bar\Delta^\vee_+$ will signify the
sets of roots, coroots, positive roots and positive coroots, respectively, of
$X_r$. The (co)root lattices will be denoted as follows:
$Q:=\Z\Pi$, $\bar Q:=\Z\bar\Pi$, $Q^\vee:=\Z\Pi^\vee$, $\bar
Q^\vee:=\Z\bar\Pi^\vee$. The weight lattices will be $P:=\Z\Psi$, $\bar
P:=\Z\bar\Psi$, and the sets of positive weights $P_+:=\Z_+\Psi$, $\bar
P_+:=\Z_+\bar\Psi$. 

\subsec{Integrable representations}

The integrable representations are unitary, highest-weight representations, for
level $k\in \Z_+$. For fixed $X_{r,k}$, the set of integrable highest weights is 
\eqn\Pkplus{P_+^k\ =\ \{\ \la=\sum_{j=0}^r\,\la_j\Lambda_j\ :\ \la_j\in \Z_+,\ 
\sum_{i=0}^r\,\la_j a^\vee_j=k\ \, \}\ ,}
where $a^\vee_j$ indicates the $j$-th comark. Let
$K=\sum_{j=0}^r\,a^\vee_j\al^\vee_j$ be the canonical central element. Then we
can rewrite  \eqn\Pkpimp{P_+^k\ =\ \{\ \la\ :\ (\al|\la)\in \Z_+, \forall\al\in
\Pi^\vee;\  (K|\la)=k\ \, \}\ .} 

A Weyl
reflection $r_\al$ normal to the root $\al$ has action
$r_\al\la=\la-(\al^\vee|\la)\al$ on weight $\la$. The primitive reflections are
then just $r_i:=r_{\al_i}=r_{\al_i^\vee}$, and they generate the affine Weyl
group $W=<r_i\,:\,i=0,1,\ldots,r>=<r_\al\ :\ \al\in \Pi^\vee>$.

The set of integrable weights $P_+^k$, or its horizontal projection $\bar
P_+^k={\cal P}(P_+^k)$ rather, can be pictured as follows. The set lies in an
alcove of finite volume in the horizontal weight space $\bar P$. The elements
of the affine Weyl group $W$ carry this alcove into copies that fill out $\bar
P$ completely. The example of $X_r=A_2$ is shown in Figure 1. 

There the
various sectors are labelled by the affine Weyl group element that carries
them into the dominant sector, the sector labelled by $id$. The elements are
written as products of primitive affine reflections $r_i$. The first 
(i.e., rightmost)
reflection is $r_0$ in all the expressions (except $id$), a choice we can
always make. The lengths of these decompositions are minimal, so that the
expressions are reduced decompositions. The number of factor primitive
reflections is then the length of the corresponding element. The thicker lines
bound the region containing the weights of $\bar P_+$. This picture will be
useful later. 

\midinsert
\vskip.25cm
\epsfxsize=6cm
\centerline{\epsfbox{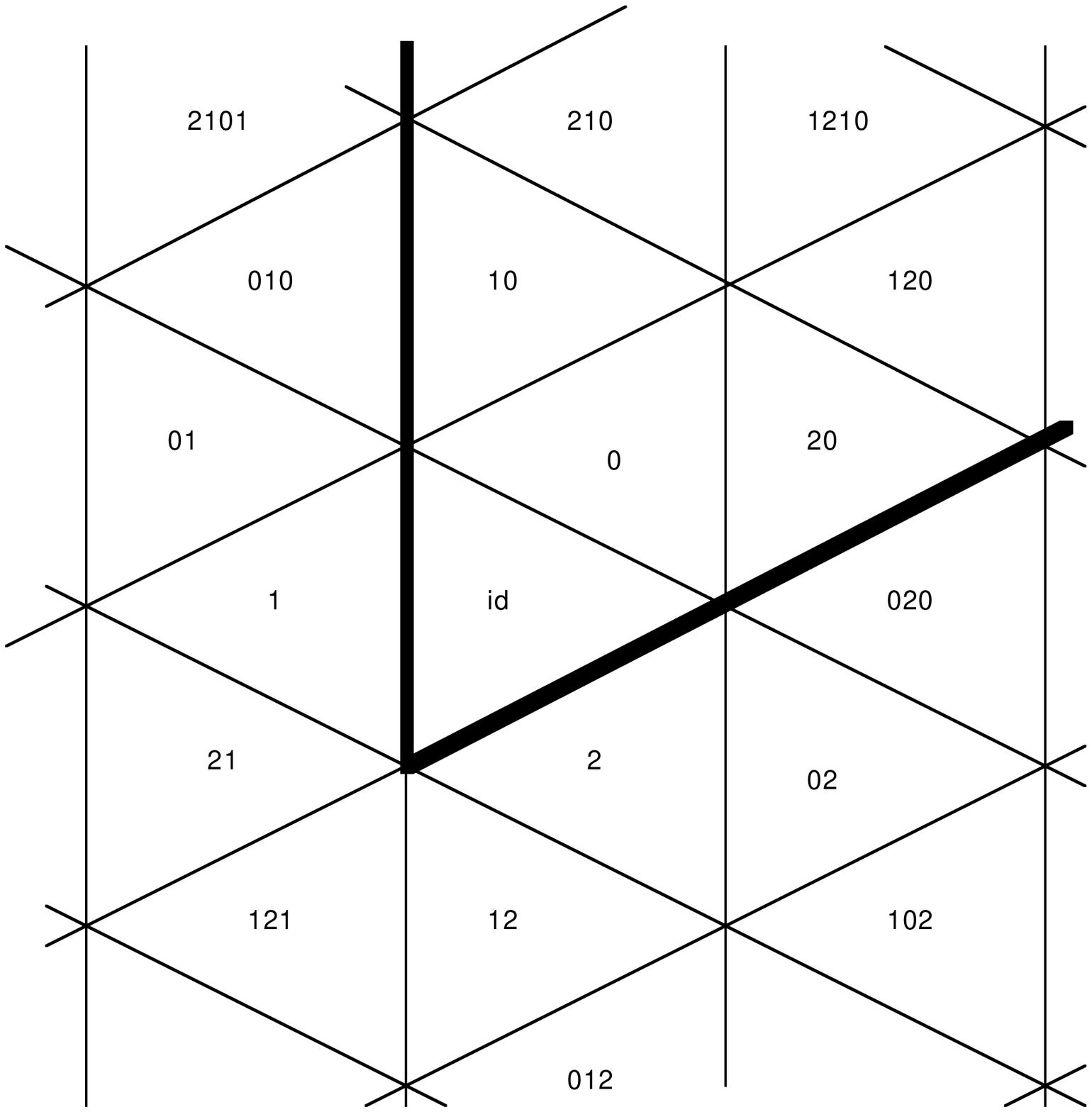}}
\vskip1cm
\leftskip=1cm
\rightskip=1cm
\noindent
\baselineskip=12pt
{\it Figure 1}.  The sectors of the weight space of the horizontal
subalgebra $A_2\subset A_{2,k}$, labelled by the affine Weyl group elements
$y\in W$ that carry them into the dominant
($y=id$) sector. The thick lines bound the
region of weights with positive horizontal Dynkin
labels. The notation $ij\cdots k$ is used
for the product $r_ir_j\cdots r_k$.\bigskip\leftskip=0cm\rightskip=0cm
\baselineskip=15pt 
\endinsert 

\subsec{Characters 
and modular transformations}

The formal character of an integrable representation $L(\la)$
of highest weight $\la$ is defined as 
\eqn\ch{{\rm ch}_\la\ =\ \sum_{\sigma\in P}\, {\rm
mult}_\la(\sigma)\, e^\sigma\ \ ,} 
where ${\rm mult}_\la(\sigma)$ is the multiplicity of weight $\sigma$ in
$L(\la)$. The elegant 
Weyl-Kac formula for this character is \eqn\wk{{\rm
ch}_\la\ =\ {{\sum_{w\in W}\,\det(w)\,e^{w.\la}} \over{\prod_{\alpha\in
\Delta_+}\,(1-e^{-\al})^{{\rm mult}(\al)}}}\ .}  Here ${\rm mult}(\al)$ is the
multiplicity of the affine root $\alpha$, and $w.\la:=w(\la+\rho)-\rho$ is
the shifted action of $w$. The normalised character is  
\eqn\normch{\chi_\la\ :=\ e^{-\delta(h_\la-c/24)}\,{\rm ch}_\la\ ,}
where $h_\la=(\la|\la+2\rho)/2(k+h^\vee)$ is the conformal weight of the
representation, and $c=k\dim X_r/(k+h^\vee)$ is the central charge. 
$h^\vee$ stands for the dual Coxeter number: $h^\vee:= (K|\rho) =\sum_{i=0}^r
a_i^\vee$. 

Suppose that a weight $\sigma$ has imaginary part $-n\delta$. Define 
\eqn\evale{e^\sigma(\tau,z,t)\ :=\ 
\exp\left[2\pi
i\left(n\tau\,+\,(\sigma|z)\,
+\,kt\right)\right]\ ,} 
where $z$ in an element of the Cartan subalgebra of the horizontal subalgebra
$X_r$. Then the normalised characters $\chi_\la(\tau,z,t)$ are the conformal
blocks for the torus partition function of the WZW conformal
field theory. Consequently, they transform nicely under the action of the torus
modular group, $PSL(2,\Z)$. This modular group is generated by $S:\
\tau\rightarrow-1/\tau$ and $T:\ \tau\rightarrow\tau+1$. Kac and Peterson found
\eqn\Schi{\chi_\la\left({-1\over\tau},{z\over\tau},{{t-(z|z)}\over{2\tau}}
\right)\
=\ \sum_{\mu\in P_+^k}\, S_{\la,\mu}^{(k)}\, \chi_\mu(\tau,z,t)\ }
and
\eqn\Tchi{\chi_\la({\tau+1},z,t)\ =\
\sum_{\mu\in P_+^k}\, T_{\la,\mu}^{(k)}\, \chi_\mu(\tau,z,t)\ .} 
The $S_{\la,\mu}^{(k)}$ ($T_{\la,\mu}^{(k)}$) are the elements of a unitary, 
symmetric 
matrix $S^{(k)}$ ($T^{(k)}$). Their precise form will be written below. 

Another expression for the normalised character is
\eqn\nchtr{\chi_\la(\tau,0,0)\ =\ e^{-2\pi
i\tau (h_\la-c/24)}\,\Tr_{L(\la)}\,e^{-2\pi i\tau L_0}\ \ ,} where $L_0$ is the
zero-mode of the Sugawara stress tensor. In \kwone, Kac
and Wakimoto defined the concept of a modular-invariant representation (of a 
complex Lie algebra $g$), by isolating the features of this last expression 
that result in the nice modular properties above. Suppose 
there exists a fixed element $E$ of a complex Lie algebra $g$ that is
diagonalisable in a representation $V$ of $g$, so that $V=\oplus_{{\cal
E}\in{\rm Spec}E}\,V_{\cal E}$, with $V_{\cal E}$ a finite-dimensional fixed
$E$-eigenvalue ($=\cal E$) subspace of $V$. If, furthermore, for some complex
number $a$,  \eqn\trE{e^{-2\pi
i\tau a}\,\Tr_V\,e^{-2\pi i\tau E}\ }
is a  modular function in $\tau$ for some principal congruence subgroup of
$PSL(2,\Z)$, then $V$ is a modular-invariant representation. 

The integrable representations are clearly modular-invariant representations
of $X_{r,k}$, with $E=L_0$, and $a=h_\la-c/24$. 
The subspaces $V_{\cal E}$ are the horizontal subspaces of
the affine representation at fixed $L_0$ eigenvalue. They represent the
horizontal subalgebra $X_r$, and because the highest weight has non-negative
integer Dynkin labels, the representations are indeed finite-dimensional.

\subsec{Admissible representations}

But Kac and Wakimoto \kwone\ found other highest-weight
representations of $X_{r,k}$ that are  modular invariant. The general class
they found, which includes the integrable representations, they dubbed the
admissible representations. They have levels that are rational, in general:
$(K|\la)=k=t/u$, with $u\in\N$, $t\in\Z$ and $\gcd(t,u)=1$. Consequently, they
have some Dynkin labels that are fractional (i.e. rational but not integer),
when the level is not integer. In particular, there are admissible
representations with highest weights having some horizontal Dynkin indices
that are not integer. This means that the horizontal subspaces of these
representations are not finite-dimensional, so that $L_0$ can not be
identified with the energy operator $E$ appropriate to them. It is a curious
feature of these admissible representations that the operator $E$ differs for
different representations, in contrast with the special case of the integrable
representations. 

Kac and Wakimoto also derived a
generalisation of the Weyl-Kac formula that applies to admissible
representations: \eqn\wkw{{\rm
ch}_\la\ =\ {{\sum_{w\in W^\la}\,\det(w)\,e^{w.\la}} \over{\prod_{\alpha\in
\Delta_+}\,(1-e^{-\al})^{{\rm mult}(\al)}}}\ ,}
which should perhaps be called the Weyl-Kac-Wakimoto formula. In it the affine
Weyl group $W$ of the Weyl-Kac formula is replaced by the group $W^\la$. The
associated Weyl group $W^\la$ of the admissible weight $\la$ is determined by
those real coroots that have integer inner product with $\la$: \eqn\Wla{W^\la\
:=\ <\,r_\al\ :\ \al\in\Delta^{\vee\,re}_+,\ (\al|\la)\in\Z\,>\ .}
 
\subsec{Principal admissible representations}

We will restrict
here to the subclass of the admissible representations that are known as
the principal admissible representations \kwtwo. In that case, the level
$k=t/u$ must be principal admissible, meaning it satisfies
\eqn\uuu{ {\rm gcd} (u,r^\vee)=1\quad , \qquad r^\vee:= {\rm
max}_i\{a_i/a_i^\vee\} }
and 
\eqn\padk{u(k+h^\vee)\ =:\ k^0+h^\vee\ ,\ k^0\in\Z_+,}
so that $k^0$ is an integrable level. The reason for this condition will
become clear later. We also have
$W^\la\cong W$, for all principal admissible highest weights, although the
action of $W^\la$ on a weight $\la$ is certainly different from the action of
$W$.   

To introduce the principal admissible highest weights, first notice that the
description of integrable highest weights begins with the set \scroots\ of
simple coroots. Define the affine Weyl vector using the dual of $\Pi^\vee$: 
\eqn\arho{\rho\ :=\ \sum_{\Lambda\in\Psi}\,\Lambda\ =\
\sum_{i=0}^r\,\Lambda_i\ ,} and let the shifted highest weight be $\la+\rho$. 
Let $v_\la$ denote the highest-weight state in the Verma
module of highest weight $\la$. The integrable condition
$(\al|\la+\rho)\in \N$ guarantees the existence of a primitive null vector  
\eqn\pnull{f_{\al^\vee}^{(\al|\la+\rho)}\,v_\la} in the Verma
module, for each $\al\in
\Pi^\vee$. (Here $f_\al$ is the lowering operator corresponding to the
positive root $\alpha$.) This leads to the characters described by \wk,\normch,
and so to their modular properties. 

A simple modification of $\Pi^\vee$ can be taken as the starting point for 
the more general class of principal admissible representations \kwtwo.
Consider
\eqn\sas{\Pi^\vee_{[u]}\ :=\ \{u\delta-\theta,
\al^\vee_1,\ldots,\al^\vee_r\, \}\ \
,} 
so that $\Pi^\vee_{[1]}=\Pi^\vee$. Among the principal admissible
representations are those whose highest weights have the same relation to
$\Pi^\vee_{[u]}$ that the integrable highest weights have to $\Pi^\vee$. 

To make this precise, we first rewrite the definition \Pkplus,\Pkpimp\ of the
set of integrable weights at level $k$ as
\eqn\Pkp{P_+^k\ =\ \{\ \la\ :\ (\al|\la+\rho)\in \N, \forall\al\in
\Pi^\vee;\  (K|\la+\rho)=k+h^\vee\ \, \}\ .}
This formula generalises to the principal admissible highest weights
in the following way. 

First, consider the analogues of $K$ and $\rho$, defined in terms of the new
set of coroots \sas. The dual of $\Pi_{[u]}^\vee$ is  \eqn\Psiu{\eqalign{
(\Pi_{[u]}^\vee)^*=:\Psi_{[u]}\ = &\ \{\Lambda_0/
u,\Lambda_1-(u-1)a_1^\vee\Lambda_0/u,\ldots,\Lambda_r-(u-1)
a_r^\vee\Lambda_0/u\, \}\cr =&\ 
\{\dot\Lambda_j:=\Lambda_j-(u-1)a_j^\vee\Lambda_0/u\ :\ j=0,1,\ldots,r\, \}\ .}}
Then    \eqn\urho{\dot\rho\ :=\ \sum_{\Lambda\in\Psi_{[u]}}\,\Lambda\ =\ 
\rho-(u-1)h^\vee\Lambda_0/u\
.} If we define $\dot\al_0^\vee:=u\delta-\theta$ and
$\dot\al_j^\vee:=\al_j^\vee$ for all $j\not=0$, so that 
\eqn\sasu{\Pi^\vee_{[u]}\ :=\ \{\,\dot\al_j^\vee\ :\ j=0,1,\ldots,r\,\, \}\ \ ,}
we can imitate $K=\sum_{i=0}^ra^\vee_i\al^\vee_i$,
and write \eqn\Ku{\dot K\ :=\ \sum_{i=0}^r\,a^\vee_i\,\dot\al^\vee_i\ \ .}

A subset $P_{u,id}^k$ of the set $P^k$ of principal admissible highest
weights is then simply 
\eqn\Pkuid{P_{u,id}^k\ =\ \{\ \la\ :\ (\dot\al|\la+\dot\rho)\in \N,
\ \forall\al\in \Pi_{[u]}^\vee;\  (\dot K|\la+\dot\rho)=k^0+h^\vee\ \, \}\ .}
The notation $P_{u,id}^k$ will be explained shortly. Here $k^0$ is a
non-negative integer, i.e. an integrable level, by \padk. So, with respect to
the system \sas, the principal admissible highest weights of $P_{u,id}^k$ are 
integrable! 

But $P_{u,id}^k$ is only a subset of the set $P^k$ of principal admissible
weights at a fixed level $k$. The important criterion turns out to be that
$\Pi_{[u]}^\vee$ is contained in the set of positive real coroots
$\Delta_+^{\vee\,re}$, since $\dot\alpha_0^\vee=(u-1)K+\alpha^\vee_0$. This
generalises to the condition 
\eqn\tyPiu{\tilde y\,\Pi_{[u]}^\vee\ \in\ \Delta_+^{\vee\,re}\ ,}
where $\tilde y$ is an element of the enlarged affine Weyl group $\tilde W$, 
that leaves invariant the set of real coroots
$\Delta^{\vee\,re}$. That is, there is a set of elements $\tilde
Y_{[u]}\in\tilde W$ one can choose such that every principal admissible weight
$\la\in P^k$ is related to $y\Pi^\vee_{[u]}$, in the way just outlined for
$\tilde y=id$. 
 The choice of set $\tilde Y_{[u]}$ is not unique, but the sets 
 $P^k_{u,\tilde y}$ of weights 
 related to the so-called simple admissible sets $\tilde y
 \Pi^\vee_{[u]}$ are disjoint for different $\tilde y$. 
In summary, we have
\eqn\PkPuy{P^k\ =\ \cup_{\tilde y\in \tilde Y_{[u]}}\,P^k_{u,\tilde y}\ .} We
can now rewrite \Pkuid\ and its generalisation in the form given by Kac and
Wakimoto:
\eqn\Pkuy{P^k_{u,\tilde y}\ =\
\{\,\la\,=\,\tilde y.\left(\la^0-(u-1)(k+h^\vee)\Lambda_0\right)\ :\ 
\la^0\in P^{k^0}_+\,\, \}\ .}
Of course, in order for the notation $P^{k^0}_+$ to make sense, we require
\padk, i.e. $k^0=u(k+h^\vee)-h^\vee\in \Z_+$.

Roughly speaking then, principal admissible highest weights are integrable, but
with respect to simple admissible sets $\tilde y\Pi^\vee_{[u]}$, rather than
$\Pi^\vee$ itself. The most important property of a principal
admissible set is that the Cartan matrix it defines is
isomorphic to that of $X_{r,k}$ (i.e. that of $\Pi^\vee$). This
implies, among other things, that the Weyl group $W^\lambda$ associated
to the principal admissible weight $\lambda$ is isomorphic to $W$. 

Furthermore, the
simplest of the simple admissible sets,
 $\Pi^\vee_{[u]}$ (i.e. the case $\tilde
y=id$), differs from $\Pi^\vee$ only in that the imaginary part $\delta$ of the
affine simple root $\alpha_0$ gets replaced by $u\delta$. Now, an affine root
with imaginary part $u\delta$ corresponds to a current with mode number $u$.
So, the replacement $\delta\rightarrow u\delta$ corresponds to a multiplication
of the mode number by $u$. The affine algebra has as a so-called winding
subalgebra \kwthree, the algebra generated by the current modes ($J_n$, say)
with mode numbers ($n$) divisible by $u$. So, we can say that principle
admissible highest weights are integrable weights for a winding subalgebra, up
to transformation by an element of $\tilde Y_{[u]}$.  

\subsec{The enlarged affine Weyl group}

The specification of the set of principal admissible weights $P^k$ boils
down to a description of the set of enlarged Weyl group elements $\tilde
Y_{[u]}$. We will give such a description in detail, but first we need to
discuss the enlarged (affine) Weyl group $\tilde W$ itself. 

The group $\tilde W$ can be described in (at least) two different ways. One
will prove useful in the next section, when we describe the admissible null
vectors, and the other is better suited to the treatment of coset field
identifications of section 4. 

First, define the group $\tilde W_+$ containing elements $\{w_j\,:\,a_j=1\, \}$,
with action $w_j\alpha_0=\alpha_j$. Its action
is similar on the fundamental weights, $w_j\Lambda_0=\Lambda_j$, and the
sets of principal admissible weights at fixed level are $\tilde W_+$-invariant:
$\tilde W_+P^k=P^k$. This is a subgroup of the group of diagram automorphisms of
the affine Coxeter-Dynkin diagrams. It is isomorphic to 
$\bar P/\bar Q$.
For $A_{r,k}$, e.g., we have $\tilde W_+\cong \Z_{r+1}$, and the relation $w_i
r_j w_i^{-1} = r_{i+j({\rm mod}\,r+1)}$ holds, where $r_j$ is any affine
primitive reflection, and $w_i^{-1}=w_{r+1-i}$. The enlarged affine Weyl group
is then \eqn\tWtWpW{\tilde W\ =\ \tilde W_+ \ltimes W\ .} 
That is, every element $\tilde y\in\tilde W$ can be written as $\tilde y=
\tilde y_+\,y$, with $\tilde y_+\in \tilde W_+$ and $y\in W$. With a
similar expression for $\tilde x\in \tilde W$, we get 
\eqn\sdprod{\tilde x\,\tilde y\ =\ \left(\tilde x_+\tilde y_+\right)\,
\left[(\tilde y_+^{-1}x\tilde y_+)y\right]\ .} 

A second description of $\tilde W$ is similar to the description of $W$ as the
semidirect product of the finite Weyl group $\bar W$, and a translation group.
Define \eqn\tbe{t_\beta\lambda\ :=\ \la+(\la|\delta)\beta - 
\left[(\la|\beta)+{1\over 2}(\beta|\beta)(\la|\delta)\right]\delta\ .} Then
the affine Weyl group is $W=t_{\bar Q^\vee}\rtimes\bar W$, where the first
factor indicates the group of translations $\{t_\beta\,:\,\beta\in \bar
Q^\vee\, \}$. The enlarged affine Weyl group is instead 
\eqn\tWtPvbW{\tilde W\ =\ t_{\bar P^\vee}\rtimes \bar W\ .} 
This means we can
 write $\tilde y=t_{\beta}\bar y$ with $\beta\in {\bar P^\vee}$ and
 $\bar y\in \bar W$, for every $\tilde y\in \tilde W$. The product of two
 elements $\tilde x= t_\gamma {\bar x},\, \tilde y= t_\beta{\bar y}\in \tilde W$
has the decomposition 
 \eqn\prodsd{\tilde x\,\tilde y\ =\ \left(t_{\gamma}t_{\bar x\beta\bar
 x^{-1}}\right)\,(\bar x\bar y)\ .}

\subsec{The set $\tilde Y_{[u]}$: a first description}

Now, to describe (a choice of) the set $\tilde Y_{[u]}$ of elements of 
$\tilde W$ required
for the specification \Pkuy\ of the set of principal admissible weights, the
important condition is \tyPiu. First we use the description \tWtWpW. 
Clearly, if
$y\in W$ satisfies this criterion, then so does $\tilde y=\tilde y_+ y$, for
any $\tilde y_+\in \tilde W_+$. So, the important task is to describe the set
of required affine Weyl group elements $Y_{[u]}$, such that $\tilde
Y_{[u]}=\tilde W_+Y_{[u]}$. 

Now \tyPiu\ with $\tilde y\rightarrow y$ implies 
\eqn\tyPiui{(\Lambda_i|y\dot\alpha_j^\vee)\ =\ 
(y^{-1}\Lambda_i|\dot\alpha_j^\vee)\ \in\ \Z_+\ ,\ }
for all $i,j\in \{0,1,\ldots,r\, \}$. First examine the case $j\not=0$. Then we
get
\eqn\yjnz{(y^{-1}\Lambda_i|\alpha^\vee_j)\ \in\ \Z_+,\ \forall
j\in\{1,\ldots,r\, \}\ ,} since $\dot\alpha_j^\vee=\alpha_j^\vee$, for $j\not=0$.
Now, the weights $\{\Lambda_i\,:\,i=0,1,\ldots,r\, \}$ are contained in the set of
integrable highest weights $P_+^a$, where $a={\rm
max}\; \{a_i^\vee\,:\,i=0,1,\ldots,r\, \}$. The horizontal projection of this set
of weights lies in an alcove of horizontal weight space, as Figure 1
illustrates for the case $X_r=A_2$. 
In that case, \yjnz\
says that only those elements $y\in W$ that label sectors in $\bar P$ 
(bounded by
the dark lines of Figure 1) need be considered. In other words, we can
restrict to consideration of $y\in W/\bar W$ only.

With $j=0$, \tyPiui\ gives
$(y^{-1}\Lambda_i|\dot\alpha^\vee_0)\in \Z_+$. Since
$\dot\alpha_0^\vee=u\delta-\theta$, this translates into 
\eqn\yjz{(y^{-1}\Lambda_i|\theta)\ \leq\ u,\ \forall
i\in\{0,1,\ldots,r\, \}\ .}
For the case $X_r=A_2$, this means that the infinite cone of triangular sectors
(labelled by elements $y\in W$) of Figure 1 is truncated to the large
triangle containing $u^2$ smaller triangular sectors. This is illustrated in
Figure 2, for the case $u=4$.

\midinsert
\vskip.25cm
\epsfxsize=7cm
\centerline{\epsfbox{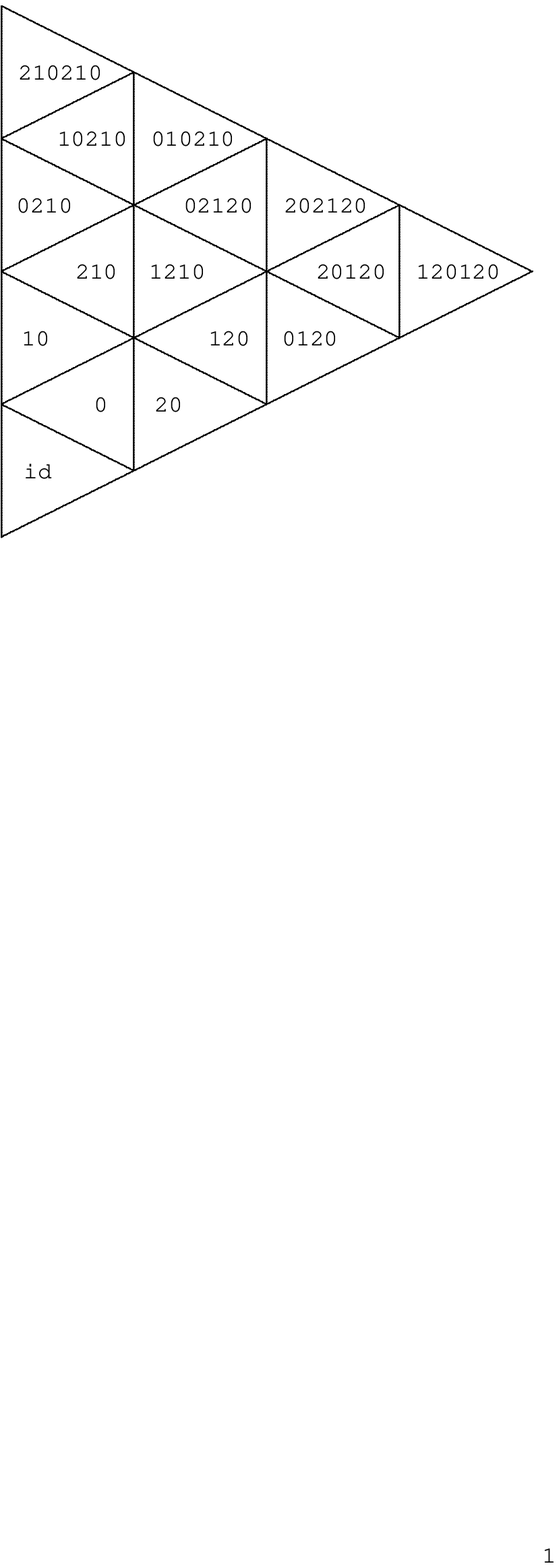}}
\vskip-11cm
\leftskip=1cm
\rightskip=1cm
\noindent
\baselineskip=12pt
{\it Figure 2}. The elements of $Y_{[4]}$
for the algebra $A_{2,k}$ (where $k=t/u$).
They are indicated, as in Figure 1, by the
sectors of $\bar P$ they carry into the
dominant sector. \bigskip\leftskip=0cm\rightskip=0cm \baselineskip=15pt 
\endinsert 

As an aid to understanding how to generate the list $Y_{[4]}$ for $A_2$,  
given in
Figure 2, we redraw the list as a graph in Figure 3. The nodes are labelled by
reduced decompositions of the elements $y$, and the lines indicate the action
of primitive reflections. The arrows on the lines give the directions
of increasing length. 
Notice that three distinct arrows arise at every node, and each node
is surrounded by three hexagons. The three types of hexagons encode
the three distinct relations: \eqn\frfr{(r_0r_1)^3=(r_0r_2)^3=(r_1r_2)^3=1\ \
.} It is now easy to see how to produce $Y_{[4]}$ for the case of $X_r=A_2$. One
need only start with $y=id$, and add to the list by multiplying by
simple reflections, such that the new elements are longer (up to a maximum
length $\ell(y)=6$)  while ensuring the elements are included in the horizontal
dominant sector. For $A_2$, we find  ${\rm card}\,Y_{[u]}=u^2$, and  $2(u-1)$
as the maximum length of an element. 

The length constraint provides a simple algebraic way to generate the
elements of $Y_{[u]}$ for $A_2$. One starts with
$y=id$, and adds  elements of increasing length, as long as one takes
care to include only those in the horizontal dominant sector, i.e. only those $y$
that are representatives of $W/\bar W$. There are different ways to characterise
these `dominant elements'. For example, they satisfy the constraint
\eqn\lcond{\ell(yr_j)\,\not<\,\ell(y)\ \ {\rm for\ }j=1,2\ .}
Equivalently, we can choose representatives of the coset $W/\bar W$ to have
reduced decompositions $y=y'r_0$ for some $y'\in W$. Any $y$ that cannot be so
written is not considered. But some that can be so written should also be
discarded: if there exists another reduced decomposition of such a  $y$ with
``first factor'' not $r_0$, it should be discarded. 
For
instance, since $r_0r_2r_0=r_2r_0r_2$, it should not be included in $Y_{[u]}$,
since it is equivalent to $(r_2r_0r_2)r_2=r_2r_0$ in the coset $W/\bar W$. 
 The elements of $Y_{[5]}$ that are not in
$Y_{[4]}$ are easily generated from the algebraic prescription:
\eqn\trort{
\eqalign{A_2: \quad  Y_{[5]}\backslash Y_{[4]} = \{ &0120120,
0210210,1201210,2101210,20120120,\cr
& 10120120,10210210,20210210,21201210\, \}\ \ .\cr} } 

\midinsert
\vskip.25cm
\epsfxsize=7cm
\centerline{\epsfbox{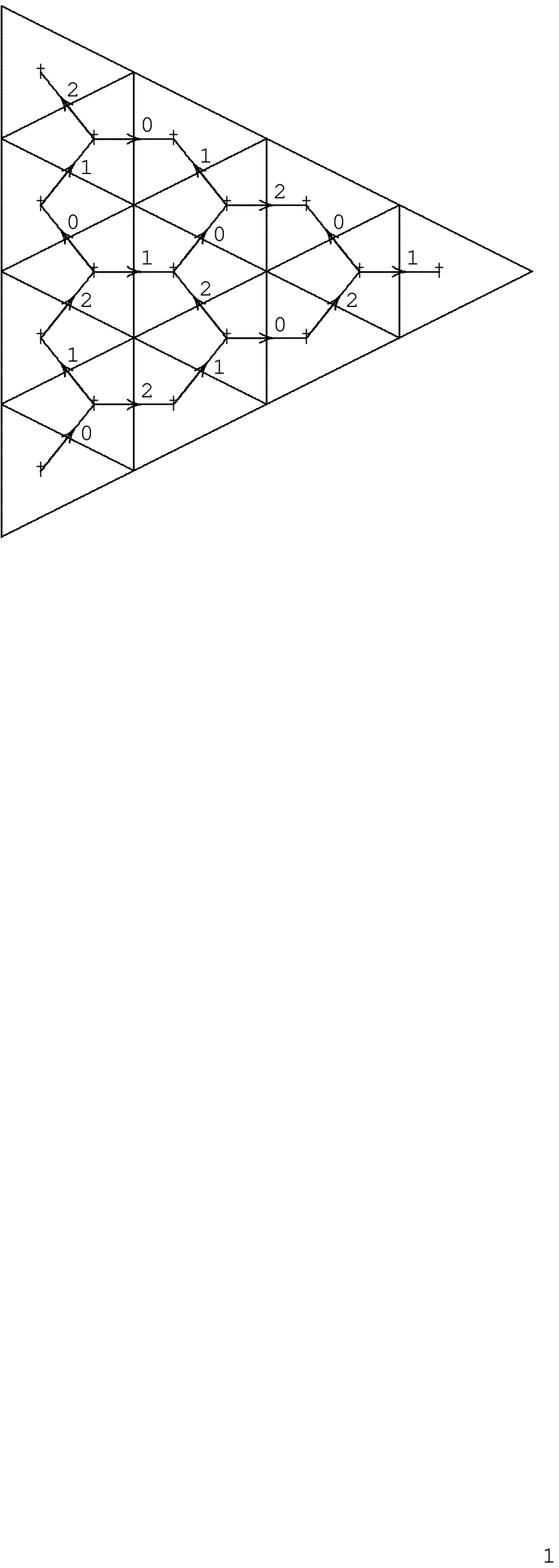}}
\vskip-11cm
\leftskip=1cm
\rightskip=1cm
\noindent
\baselineskip=12pt
{\it Figure 3}. A graph indicating the
elements of $Y_{[4]}$ for $A_2$. The arrows
on the lines indicate the directions of
increasing length. \bigskip\leftskip=0cm\rightskip=0cm \baselineskip=15pt 
\endinsert 

A simple length constraint does not hold in general but a modified one also
applies to the
$A_1$ case, for which  $Y_{[u]}$ 
 is given by all the elements of $W$ that start with $r_0$ and have 
$\ell(y)\leq u-1$, e.g. 
\eqn\asas{A_1: \quad  Y_{[5]}=
\{id,0,10,010,1010\, \}}

For the other algebras, no simple characterization has been found.  Notice
however that once the condition  $y\in W/\bar W$, or equivalently
\eqn\lcondr{\ell(yr_j)\,\not<\,\ell(y)\ \ {\rm for\ }j=1,2, \ldots , r\  ,}
has been taken into account, \tyPiu\ (with $\tilde y\rightarrow y$) reduces to
the single condition \eqn\ssd{y\, [u\al_0^\vee+(u-1)(\al_1^\vee+\cdots +
\al_r^\vee)]\,=\,y\alpha_0^\vee+(u-1)K\,\in\ \Delta_+^{\vee\,re} \, , }
where we have used $yK=K$. 
In this way it is easy to generate for instance the set
$Y_{[2]}$ for
$A_3$:
\eqn\tats 
{ A_3: \quad Y_{[2]}= \{id,0,10,30,210,230,310,0310\, \}\ \ .}  The cases $C_2$
and $G_2$ are treated in an Appendix.

As mentioned above,
we will otherwise restrict consideration to the algebras $A_{r,k}$. For them, 
all admissible highest weights are principal admissible. For other algebras,
however, there are admissible highest weights that are not also principal
admissible.

\subsec{The set $\tilde Y_{[u]}$: a second description}

A second description of the set $\tilde Y_{[u]}$ is obtained using
\tWtPvbW. 
With $\tilde y=t_\beta\bar y$, $\beta$ turns out to lie in the coset $\bar
P^\vee/u\bar Q^\vee$ centered around the origin, that is, in
\eqn\Cb{\bar C\ :=\ \{\,\beta\in \bar P^\vee
=\bar P\ :\ -u< (\beta|\alpha)\leq u ,\ \forall\ 
\alpha\in \bar \Delta_+^\vee\,\}\ .} 
Notice that this
 result determines   \eqn\bvol{{\rm card}\,\tilde Y_{[u]}\ =\ 
 \vert\bar P^\vee/u\bar Q^\vee\vert =\  u^r\,\vert \bar P/\bar Q\vert\ =\ 
 u^r\,\vert\tilde W_+\vert\ .}
Given a value of $\beta$ in this set, there is a unique $\bar y$ that ensures
\tyPiu\ (as proved in \kwtwo).  Actually, this $\bar y$ is the shortest
element of $\bar W$ such that 
 \eqn\ybcondi{(\bar y^{-1}\beta|\alpha^\vee_j)\le 0,\ {\rm for}\ j=1,\ldots,r\
.}

The subset $Y_{[u]}\subset \tilde Y_{[u]}$ is obtained by restricting $\beta$ to
be an element of ${\bar Q}^\vee$.

To illustrate this second
description, consider the  elements of
$Y_{[3]}$ for
$A_2$:
\eqn\sdsd{\eqalignT{
&id= t_{(0,0)}\qquad  &r_0=t_{(1,1)}\, r_1r_2r_1\qquad 
&r_1r_0=t_{(-1,2)}\,r_2r_1\cr
& r_2r_0=t_{(2,-1)}\, r_1r_2\qquad
& r_1r_2r_0=t_{(-2,1)}\, r_2\qquad 
&r_2r_1r_0=t_{(1,-2)}\, r_1\cr
&r_0r_2 r_1r_0=t_{(3,0)}\, r_1r_2\qquad  &r_0r_1 r_2r_0=t_{(0,3)}\, r_2r_1\qquad
&r_1r_2 r_1r_0=t_{(-1,-1)}\cr}}
Here we use the notation
$(\beta_1,\beta_2)=\beta=\beta_1\Lambda_1+\beta_2\Lambda_2$. All values of
$\beta$ appearing above span the set $(\beta_1,\beta_2)$ such that
\eqn\dr{-3<\,\beta_1,\beta_2, \beta_1+\beta_2\,\leq 3 \; .}
with the additional
constraint that $\beta_1+2\beta_2= 0 $ mod $3$ (ensuring that
$\beta\in{\bar Q}^\vee$).  
The above list of elements $t_\beta \bar y$ could thus
have been derived by first specifying those $(\beta_1,\beta_2)$ that satisfy the
above conditions and fixing
$\bar y$  from the requirement that
${\bar y}^{-1}$ be the shortest element of $\bar W$ that maps a $\beta$ that
has at least one positive label to a weight with its two labels non-positive.

Most of this section is an account of our understanding of the results of Kac 
and Wakimoto \kwone\kwtwo. What is new are the detailed descriptions of how to
obtain the sets of principal admissible weights. (A different, less natural
description was discussed in \mw.) There is some overlap with the treatment in
\fgp\fgptwo, where the authors concentrate on the special case $k^0=0$.

\newsec{Malikov-Feigen-Fuchs Null Vectors}

The existence of null vectors in a Verma module is responsible for the Weyl
symmetry of the weight diagrams of the corresponding integrable highest-weight
module. For example, an $A_1$ highest weight $\la=\la_1\Lambda_1$ produces a
Verma null vector $f_{1}^{\la_1+1}v_\la$ if $\la_1\in \Z_+$, where we use
the notation $f_1=f_{\al_1}$. Then the corresponding integrable highest weight
representation has a $\Z_2$ Weyl symmetry, generated by the reflection that
takes the highest weight $\la$ to the lowest weight. If $\la_1\not\in \Z$,
however, there are no Verma null vectors and so no lowest weight; the
highest-weight module and Verma module coincide. The highest-weight representation
is infinite-dimensional, and there is no Weyl reflection symmetry. 

The Weyl group $W^\la$ associated to a principal admissible highest weight
$\la$ is isomorphic to the affine Weyl group $W$. This means there are,
roughly speaking, just as many null
vectors in a principal admissible representation as there are in an integrable
representation of $X_{r,k}$. As already mentioned above (see \pnull), the null
vectors in a $X_{r,k}$ Verma module with integrable highest weight $\la$, are
all descendants of $r+1$ primitive null vectors, $n_j:=f_j^{\la_j+1}v_\lambda$,
with $j=0,1,\ldots,r$. So, the principal admissible repesentations should also
have $r+1$ primitive null vectors. 

Malikov, Feigen and Fuchs \ref\mff{F. Malikov, B. Feigen and 
D.B. Fuchs, {\sl Funct. Anal. Appl.} {\bf 20} (1986) 103}\ showed how to write
the null vectors of Verma modules with highest weights that are not integrable,
in an unconventional, but beautiful form. Here we point out a simple, direct
way of writing their expressions for the primitive null vectors of principal
admissible representations.  

Let $\{f_i,h_i,e_i\,:\,i=0,1,\ldots,r\, \}$ denote the Chevalley generators of
$X_{r,k}$.  So \eqn\fhe{\eqalign{[h_i,e_j]\ =\ A_{ij}\,e_j\ &,\ \ \ 
[h_i,f_j]\ =\ -A_{ij}\,f_j\ ,\cr
[e_i,f_j]\ =\ \delta_{ij}\,h_i\ &,\ \ \forall\ i,j\in\{0,1,\ldots,r\, \}\ ,
}}
where $A_{ij}=(\alpha_i|\alpha^\vee_j)$ are the elements of the Cartan matrix.
What Malikov-Feigen-Fuchs (MFF) found was that \pnull\ generalises to
\eqn\mffi{f_{i_\ell}^{x_\ell}\cdots f_{i_2}^{x_2}f_{i_1}^{x_1}\,v_\lambda\ ,}
but with the exponents $\{x_j\, \}$ complex, not necessarily integer, depending on
the highest weight $\la$. For admissible $\la$, the exponents are rational. 
Such expressions are meaningful because they can be rewritten in conventional
form, i.e. as linear combinations of expressions of  the type \mffi, but with 
non-negative integer exponents. 

The rewriting procedure is straightforward, but tedious. Identities
involving integral powers are derived, and when the coefficients that appear are
polynomial, they can be continued to complex exponents. Such
continued relations may be used to re-express the MFF ``continued monomials''
in conventional form. For a helpful simple example, see the Appendix of 
\ref\baso{M. Bauer and N. Sochen, {\sl Comm. Math. Phys.} {\bf
152} (1993) 127}. For more help, see
\ref\fgpb {P. Furlan, A.Ch. Ganchev and V.B. Petkova, {\sl Nucl. Phys}. {\bf
B431} (1994) 622.}].

First notice that the integrable null vectors \pnull\ are just 
\eqn\Fdef{F_i\,v_\la\ ,\ \ i=0,1,\dots,r;\ \ \ \ F_i:=\,{\cal
N}(f_i^{1+h_i})\ ,} where the ``normal ordering'' ${\cal N}$ simply
specifies that the 
Cartan subalgebra
operators should be moved to the right: \eqn\nord{{\cal N}(h_i^b f_i^a)\ =\
{\cal N}(f_i^a h_i^b)\  =\ f_i^a h_i^b\ .} Operators like the $F_i$ were
considered in  \ref\bota{P. Bowcock and A. Taormina, {\sl Comm. Math. Phys.}
185 (1997) 467}. 

An MFF null vector exists whenever the highest weight $\la$
satisfies $(\la+\rho|\alpha^\vee)=n\in\N$, for some real root $\alpha\in
\Delta^{\vee\,re}$. By definition, a real root $\alpha$ is Weyl equivalent to a
simple root: $\alpha=v\alpha_j$, where $\alpha_j\in\Pi$. Now, for admissible
weight $\la\in P_{u,\tilde y}^k$, we have 
\eqn\kkadm{(\la+\rho|\tilde y\dot\alpha^\vee_j)\ =\ (\tilde
y\big(\sum_{\ell=0}^r\,(\la^0_\ell+1\big)\dot\Lambda_\ell)|\tilde
y\dot\alpha^\vee_j)\ =\ \la^0_j+1\ \in\ \N\ .}
We find that the description of the element $\tilde y\in \tilde Y_{[u]}$
leads directly to an expression for the MFF null vectors, in terms of the
operators $F_j$. 

Precisely, suppose 
\eqn\yrd{\tilde y\ =\ \tilde y_+\,y\ =\ \tilde y_+\,(r_{i_\ell}\cdots
r_{i_1})\ ,}   with $\tilde y_+\in\tilde W_+$,
and $(r_{i_\ell}\cdots
r_{i_1})$ a reduced decomposition of $y\in W$. Define 
\eqn\typF{\tilde y_+\,F_j\,\tilde y^{-1}_+\ :=\ F_\ell\ ,\ \ {\rm if}\ \tilde
y_+\al_j=\al_\ell\ .} 
Then there are MFF null vectors of the form
\eqn\adnvnz{n_j\ :=\ \tilde y_+\, (F_{i_\ell}(\cdots (F_{i_1}(F_j)F_{i_1})
\cdots )F_{i_\ell})\,\tilde y_+^{-1}\,v_\la\ ,\ \ \forall j=1,\ldots,r\ .} 
The remaining null vector, corresponding to the case $j=0$,
is only a little more complicated. The difference is that while
$\dot\al^\vee_j=\al_j^\vee$ for $j\not=0$, we have
$\dot\al_0^\vee=(u-1)K+\al_0^\vee$. However, since $\dot\al_0^\vee$ is real, one
always has  $\dot\al^\vee_0=z\al^\vee_q$, for some $z\in W$, and some
$q\in\{0,1,\ldots,r\, \}$. If $yz=r_{p_s}\cdots r_{p_1}$ is a reduced
decomposition of $yz$, then we simply get
\eqn\adnvz{n_0\ :=\ \tilde y_+\, 
(F_{p_s}(\cdots(F_{p_1}(\,F_q\,)F_{p_1})\cdots)F_{p_s})\,\tilde y_+^{-1}\,v_\la
\ .} 

Another expression for the operators defined above is \eqn\Flog{F_i\ =\
\sum_{\ell=0}^\infty\,{1\over {\ell!}} ({\rm log} f_i)^\ell\,  (1+h_i)^\ell\ ,}
where we use
\eqn\logf{{\rm log}\,f_i\ =\ \sum_{n=1}^\infty\, {{-1}\over n}\,
(1-f_i)^n\ .}
With these expansions, it is straightforward to verify that 
\eqn\hiFj{[h_i,F_j]\ =\ -A_{ij}F_j(1+h_j)\ ,}
for example. This suffices to show that the expressions \adnvnz\ and
\adnvz\ reduce to those prescribed in \mff. 

For convenience, we point out that if $\al^{[u]}_0$ denotes $\dot\al_0^\vee$
at fixed $u$, then $\al_0^{[u+2]}=r_0r_\theta\,\al_0^{[u]}$. Also, for the
case of $X_r=A_r$, we have 
\eqn\rth{r_\theta\ =\ (r_1r_2\cdots r_r)(r_1\cdots r_{r-1})\cdots(r_1r_2)r_1\
,}
and $\al_0^{[2]}=r_0r_1r_2\cdots r_{r-1}\al_r^\vee$. 

As a simple example, consider the four weights $[-{1\over 2},0],$ $[-{3\over
2},1],$ $[1,-{3\over
2}],$  $[0,-{1\over 2}],$ of $P^{-{1\over 2}}$, for $A_{1,-{1\over 2}}$. The
first two can be considered elements of $P^{-{1\over 2}}_{2,id}$, the third
and fourth, elements of $P^{-{1\over 2}}_{2,w_1}$. We also need
$\dot\alpha_0^\vee=K+\alpha^\vee_0 = r_0\alpha^\vee_1$ for $u=2$. For 
$[-{1\over 2},0],$ we get the null vectors
\eqn\nxi{\eqalign{n_0\ =&\ F_0F_1F_0\,v_{[-{1\over 2},0]}\ =\ 
f_0^{7\over 2}f_1^2f_0^{1\over 2}\,v_{[-{1\over 2},0]}\ ,\cr
n_1\ =&\ F_1\,v_{[-{1\over 2},0]}\ =\ 
f_1\,v_{[-{1\over 2},0]}\ ;\cr}}
and for $[-{3\over 2},1]$
\eqn\nxii{\eqalign{n_0\ =&\ F_0F_1F_0\,v_{[-{3\over 2},1]}\ =\ 
f_0^{5\over 2}f_1^1f_0^{-{1\over 2}}\,v_{[-{3\over 2},1]}\ ,\cr
n_1\ =&\ F_1\,v_{[-{3\over 2},1]}\ =\ 
f_1^2\,v_{[-{3\over 2},1]}\ .\cr}}
The null vectors for highest weights $[1,-{3\over
2}]$ and $[0,-{1\over 2}]$ can be simply obtained from these by applying
$w_1$, which changes the highest weights and interchanges $F_0$ and $F_1$.  

A more compact notation suggests itself. Suppose $\al$  is a real root, so
that $\al=\tilde w\al_j$, for some $\tilde w\in \tilde W$, and that $\tilde
w=w_m w$ with $w\in W$ having reduced decomposition $w=r_{i_1}r_{i_2}\cdots
r_{i_\ell}$. Then define  \eqn\Fal{F(\al)\ :=\
w_m(F_{i_\ell}(\cdots(F_{i_1}(F_j)F_{i_1})\cdots)F_{i_\ell})w_m^{-1}\ .} 
It would be interesting to investigate the properties these operators. What we
have just observed is simply that the MFF null vectors take the form 
\eqn\mmfF{n_j\ =\ F(\tilde y\dot\al_j^\vee)\,v_\la\ ,} 
for principal admissible highest weights $\la\in P^k_{u,\tilde y}$. 

A general feature of the null vector \mmfF\ is that its inner exponent is just
the integer $1+\lambda^0_j$. See the $A_{1,-{1\over 2}}$ example above, for
instance, where for highest weights $\lambda = [-{1\over 2},0],$ $[-{3\over
2},1],$ $[1,-{3\over
2}],$  $[0,-{1\over 2}],$ the ``integrable parts'' are
$\lambda^0 = [1,0],\ [0,1]$, $[0,1],\ [1,0]$, respectively. Explicitly, we can
rewrite \mmfF\ as
\eqn\mmfFi{n_j\ =\
w_m(F_{i_\ell}(\cdots(F_{i_1}(f_i^{1+\lambda^0_j})F_{i_1})
\cdots)F_{i_\ell})w_m^{-1}\,v_\lambda\ ,  }
where for $j\not=0$, $i=j$ and $\tilde y=w_mr_{i_1}\cdots r_{i_\ell}$; and for
$j=0$,  
 $\dot\alpha^\vee_0=z\alpha^\vee_i$ and 
$\tilde y z=w_mr_{i_1}\cdots r_{i_\ell}$. 
This is helpful when writing the MFF expressions. For example, for $A_{2,k}$
with $u=3$ and $\lambda=k\Lambda_0$, we have $\lambda^0=k^0\Lambda_0$ and
$\dot\alpha_0^\vee = r_0r_\theta\alpha_0^\vee = r_0r_1r_2r_1\alpha_0^\vee$. So
we can write  \eqn\mffxii{\eqalign{n_0\ =&\
F_0F_1F_2F_1(f_0^{1+k^0})F_1F_2F_1F_0\,v_{k\Lambda_0} \cr
  =&\
f_0^{13+5k}f_1^{5+2k}f_2^{10+4k}f_1^{5+2k}
f_0^{7+3k} f_1^{2+k}f_2^{4+2k}f_1^{2+k}f_0^{1+k} \,v_{k\Lambda_0}\ .\cr}}

\newsec{Field Identifications in Nonunitary Coset Models}

\subsec{The problem of field identifications}

The Virasoro minimal models have a well-known  $A_1$ diagonal coset
description: 
\eqn\a{{A_{1,1}\oplus A_{1,k}\over A_{1,k+1} }\ \ .}
The level $k$ is related to the usual parameters $p,p'$ ($p, p' $ coprime and
$p>p'$) that appear in the expression of the central charge 
\eqn\ccr{ c= 1-{6(p-p')^2\over pp'}}
by the following relations:
\eqn\das{ p-p'= u\qquad, \qquad  k= {3p'-2p\over p-p'}\ \ .}
These models are unitary when $p=p'+1$ and the Virasoro fields are expressed in
terms of constituent integrable representations $\{\eta, \la; \la'\}$ of $A_1$ (at
respective levels
$1,\, k,
\, k+1$). The basic Virasoro field identification \eqn\bfbf{ \phi_{r,s} \sim
\phi_{p'-r, p-s}}
(where the labels $r,s$ runs over the range $1\leq r\leq p'-1\,,1 \leq s\leq
p-1$, i.e., $r=\la_1+1, \, s=\la_1'+1$) is simply  a translation of the coset
field identification  
\eqn\idet{ \{\eta, \la; \la'\}\sim \{A\eta, A\la; A\la'\} }
where $A$ is the outer automorphism that interchanges the roots $\al_0$ and
$\al_1$.  In coset models, field identifications generically result from
outer automorphisms \ref\MS{G. Moore and Seiberg,
{\sl Phys. Lett.} {\bf B220}
(1989) 422; D. Gepner, {\sl Phys. Lett.}
{\bf B222} (1989) 207; W. Lerche, C. Vafa and N. Warner,
{\sl Nucl. Phys.} {\bf B324} (1989) 427; C. Ahn and M.A. Walton,
{\sl Phys. Rev.} {\bf D41} (1990) 2558} (exceptions define the
so-called maverick cosets \ref\DJ{D. Dunbar and K. G. Joshi, {\sl Int. J.
Mod. Phys.} {\bf A8} (1993) 2803 and {4103}}).

{}From the Virasoro point of view, the distinction between unitary and
nonunitary theories (apart from unitarity) is rather smooth. In particular, 
there is no dramatic increase in the number of distinct fields when passing from
unitary to nonunitary models (for instance, with $p+p'$ fixed).  However, from
the coset perspective, there is quite a difference in the number of admissible
representations (two of the three representations being admissible when $p>p'+1$)
when, say,  we start with
$k= k^0$ and then  turn on the value of $u-1$ from 0 to positive
integers. Therefore, in the nonunitary case there must be a much larger number
of field identifications than those resulting from the action of the outer
automorphims.  Our goal is to unravel this whole set in a systematic way.  We 
show in particular that these identifications have
a natural formulation in terms of the  description of the principal admissible
representations presented in section 2.   

\subsec{Branching conditions}

Let us consider the diagonal coset
\eqn\a{{X_{n,\ell}\oplus X_{n,k}\over X_{n,k+\ell} }}  
where $\ell$ is a non-negative integer and $k$ is admissible.
Let $\eta$ be an integrable weight of level $\ell$,  $\la$ and $\la'$ be two
admissible weights at levels $k$ and $k+\ell$, respectively:
\eqn\fia{\la \in P^{k}_{u,\y}\ ,\qquad \la' \in P^{k+\ell}_{u,\y'}\ . }
The basic condition
for the triplet $\{\eta, \la; \la'\, \}$ to be a possible label of a coset field
is that the following branching condition be satisfied:
\eqn\bc{ {\cal P}(\gamma + \la- \la')\in {\bar Q}\ ,}
where, as above, ${\cal P}$ projects an affine weight onto its horizontal part.

More explicitly, the branching condition reads
\eqn\fiaa{ {\cal P}\left(\eta + \y. \la^0+ \y((k-k^0)\Lambda_0) -\y'.(
\la'^0)+\y'((k+\ell-k^0-u\ell)\Lambda_0)\right)\in{\bar Q}}
This requires the cancellation of the fractional parts modulo
$u{\bar Q}$, that is 
\eqn\faiab{{\cal P}({\tilde y} k\Lambda_0 - {\tilde y}'k\Lambda_0) = 0 ~
{\rm mod}~{\bar Q}\ . } This forces
\eqn\fiac{{\cal P}({\tilde y} \Lambda_0- {\tilde y}'\Lambda_0) = 0 ~{\rm mod} ~
u{\bar Q} } With
${\tilde y}=  t_\beta {\bar y} $ and given that  ${\bar y}\in\bar W$ (so that
${\bar y}\Lambda_0=\Lambda_0$) we have
\eqn\fiad{ t_\beta \Lambda_0 = t_{\beta'}\Lambda_0 + u\gamma }
 with $\gamma\in {\bar Q}$. This
implies that
\eqn\fiae{ \beta= \beta'+u\gamma\ \ . }
Since $\beta\in {\bar P}^\vee$, it is plain that $\gamma
\in{\bar Q}^\vee$. Given that $\beta$ and $\beta'$ are associated to admissible
weights, they can be taken to satisfy the condition 
\eqn\fiaf{-u<(\beta|\alpha)\leq u\; \quad {\rm and} \qquad -u<(\beta'|\alpha)\leq
u\ \ } for any positive root  $\alpha$. 
Hence the relation $\beta= \beta'+u\gamma$ can be satisfied only if 
$\gamma=0$. Therefore $\beta=\beta'$ and since  ${\bar y}$ is uniquely specified
by 
$\beta$, 
it follows that ${\bar
y}={\bar y}'$ and thus
${\tilde y}={\tilde y}'$.  This simplifies the branching condition substantially, 
reducing it to \eqn\fiag{{\cal P} \left(\eta+  \y.[\la^0-\la'^0
-\ell(1-u)\Lambda_0]\right)\in{\bar Q}\ \ .} This is clearly equivalent to
requiring \eqn\fiagg{ {\cal P}( \eta + \la^0-\la'^0)\in{\bar Q}\ \ . }

Because the Weyl transformation of any weight does not affect the branching
condition, it is clear that if  $\{\eta, \la; \la'\, \}$ labels a coset field,
$\{\eta, w.\la; w.\la'\, \}$ also labels a coset field provided that
$w.\la$ is admissible (which thereby ensures the admissibility of $
w.\la'$). (Of course there is no point in applying the Weyl transformation to
the first weight since in that way, one can never reach an integrable weight. 
Also, the above analysis shows that the same Weyl transformation must be used for
the two admissible weights.)

There is another general class of transformations of the coset field
that preserve the branching condition.  It involves the action of outer
automorphisms.  Every
 outer automorphism $A\in {\tilde W}_+$ can be put in a 1-1 correspondence with an
element of the finite Weyl group $w_A$ via the relation 
\eqn\ber{A\la= k(A-1)\Lambda_0+w_A\la}
It readily follows from this that if $\{\eta, \la; \la'\, \}$ labels a coset
field, $\{A\eta, A\la; A\la'\, \}$ also 
labels a coset field since \eqn\fiah{{\cal
P}\left(\ell(A-1)\Lambda_0+w_A\eta
+k(A-1)\Lambda_0+w_A\la-(k+\ell)(A-1)\Lambda_0+w_A\la'\right)\in {\bar Q}} 
reduces to the condition \bc.

These transformations are
candidates for field identifications, since they do not modify the branching
conditions. To demonstrate that two coset fields can be identified, we must
show that their characters are the same, or equivalently, that their modular
transformation matrices are identical. 

\subsec{Field identifications via modular matrices}

In terms of the coset components, the coset modular $S$ matrix is
\eqn\sco {S_{\{\eta,\la;\la'\, \}\{\xi,\mu;\mu'\, \}}^{{\rm coset}} = 
S^{(\ell)}_{\eta\xi} S^{(k)}_{\lambda\mu} {S^{(k+\ell)}_{\la'\mu'}}^*
}
The $S$ matrix entering on the rhs is
\eqn\smat{\eqalign{ S_{\la\mu}^{(k)} &= F_k e^{-2\pi i[(\mu^0+\rho|
\be)+(\la^0+\rho|\ga)+(k+h^\vee)(\be|\ga)]}\cr
\qquad&\qquad\qquad\qquad\times\sum_{w\in {\bar W}}
({\rm det}w) e^{-{2\pi i\over k+h^\vee} 
[(w(\la^0+\rho)|\mu^0+\rho+(k+h^\vee)\gamma)]} \cr &= 
F_k e^{-2\pi i[(\mu+\rho|
\be)]}({\rm det}{\bar x}) \sum_{w\in \bar W}
({\rm det}w) e^{-{2\pi i\over k+h^\vee} [(w(\la^0+\rho)|\mu+\rho)]}\cr
&= 
F_k e^{-2\pi i[(\la+\rho|
\gamma)]} ({\rm det}{\bar y})\sum_{w\in \bar W}
({\rm det}w)e^{-{2\pi i\over k+h^\vee} [(w(\la+\rho)|\mu^0+\rho)]}\cr} }
where $F_k$ is a constant fixed by the unitarity of $S^{(k)}$, and we used the
notation  \eqn\fiaj{\eqalign{
\la &=\y. [\la^0+(k-k^0)\Lambda_0],\qquad \y = t_\be {\bar y}\cr  \mu &= \xt.
[\mu^0+(k-k^0)\Lambda_0],\qquad \xt = t_\gamma {\bar x}\ \ . \cr}} For the
second line we used
\eqn\fiajj{(\be | w\la)\,\equiv\, (\be | \la)\qquad {(\rm mod} \; 1)\ ,}
 for any $\la$ (i.e., $\be\in {\bar P}^\vee$) and 
\eqn\fiai{\eqalign{
 \mu+\rho &= \xt. [\mu^0+(k-k^0)\Lambda_0] +\rho\cr 
 &= {\bar x}( \mu^0+\rho) +
(k^0+h^\vee)\gamma +(k-k^0)[\Lambda_0+\gamma]\cr& = {\bar x}(
\mu^0+\rho) + (k+h^\vee)\gamma +(k-k^0)\Lambda_0\ \ .\cr}}

A simple analysis yields the following two relations:
\eqn\srel{ \eqalign{ S_{w.\la, \mu}^{(k)} & = ({\rm det}w) S_{\la, \mu}^{(k)}\cr
S_{A\la, \mu}^{(k)} & =  S_{\la, \mu}^{(k)}~e^{-2\pi i(A\Lambda_0|\mu)}~
e^{-2\pi i((w_A-1)\beta|\mu+\rho)}\cr}}
The first result follows readily from the third relation in \smat.  To obtain
the second one, we use \ber\  -- notice that if
$k$ is integer, the second phase factor in it disappears. From the first of
these relations, it follows directly that 
\eqn\fiak{S_{\{\eta,w.\la;w.\la'\, \}\{\xi,\mu;\mu'\, \}}^{{\rm coset}}  =
S_{\{\eta,\la;\la'\, \}\{\xi,\mu;\mu'\, \}}^{{\rm coset}}}
The second one implies 
\eqn\fiakk{S_{\{A\eta,A\la;A\la'\, \}\{\xi,\mu;\mu'\, \}}^{{\rm coset}} = 
S_{\{\eta,\la;\la'\, \}\{\xi,\mu;\mu'\, \}}^{{\rm coset}}~\phi}
with 
\eqn\fiakl{\phi = e^{-2\pi i(A\Lambda_0|\xi+\mu-\mu')}~
e^{-2\pi i[((w_A-1)\beta|\mu+\rho)-((w_A-1)\beta'|\mu'+\rho)]}}
where $\beta'$ refers to the weight $\la'$.  But the branching condition
requires $\beta'=\beta$, the second part of the phase reduces to 
$((w_A-1)\beta, \mu-\mu')$. The branching condition for the coset field
$\{\xi,\mu;\mu'\, \}$ forces $\mu-\mu'$ to be an integer weight.  Now, because
$\beta$ is a coweight, the product of $(w-1)\beta$  for any Weyl group
element $w$ with an integer weight is necessarily an integer. 
The second phase factor is thus 1.  For the first one,
we notice that the finite part of $\xi+\mu-\mu'$ (the part that contributes
to the scalar product) is an element of $\bar Q$ and since $A\Lambda_0\in 
{\bar P}^\vee$, the scalar product  $(A\Lambda_0|\xi+\mu-\mu')$ is also an
integer. Therefore
$\phi=1$ and we have established the $S$-part of the field identifications.

The analysis of the $T$-part is much simpler since it relies solely on the
behaviour of the conformal dimensions under the transformations under
consideration.  For any conformal field theory, the matrix $T$ takes the 
following simple form
\eqn\sada{T_{ij} = e^{2\pi i (h_i-c/24)}\delta_{ij}\ ,}
where $h_i$ is the $i$th conformal weight. The conformal weight corresponding
to the $X_{r,k}$ representation of highest weight $\lambda$ is defined above,
after \normch.  

It is not simple to find an expression for the 
conformal dimension of the coset field $\{\eta, \la; \la'\, \}$ in terms of
those of its ``components''. But it can only differ by an integer  from the 
linear combination: \eqn\fiall{\Delta h = h_\eta + h_\la-h_{\la'}\ \ .} This is
easily checked to be invariant under the two field transformations considered
above (modulo an  integer in the second case).  We omit the details.

The $S$ and $T$ invariance proves
then  the coset field identifications
\msw
\eqn\fiam{  \{\eta, \la; \la'\, \}\sim
\{\eta, w.\la; w.\la'\, \}\sim \{A\eta, A\la; A\la'\, \}\sim
\{A\eta,Aw.\la; Aw.\la'\, \}\ .}

\subsec{Coset representatives}

Let us now consider a chain of field identifications that will lead
to a canonical choice of coset field representatives. Let us start with
admissible fields corresponding to the element $\y=id$:
\eqn\fiann{\la\equiv \La \in P^{k}_{u,id}, \qquad \la'\equiv \La' \in
P^{k+\ell}_{u,id}}
by picking up two weights $\la^0\in P^{k^0}, \, \la'^0\in P^{k^0+\ell
u}$ together with a coset partner $\eta \in P^{\ell}$ such that ${\cal
P}(\eta+\la^0-\la'^0)\in{\bar Q}$. We now construct a string of $w$-type
field identifications, by choosing $w\in Y_{[u]}$ (the subset of affine and finite
Weyl transformations among the whole set $\Y_{[u]}$). This ensures that both
$w.
\La$ and
$w.
\La'$ are admissible.  That yields then $|Y_{[u]}|$ field
identifications: $\{\eta,\La;\La'\, \}\sim\{\eta,w.\La;w.\La'\, \}$. 
On each coset field, we then act on the three weights with all possible
elements
$A$ of the outer automorphism group.  This produces $|\W_+|$
additional field identifications. Hence, starting from a
coset field with $\y=id$, we generate $|\Y_{[u]}|$ field 
identifications. This is
necessarily a complete set of field identifications. Indeed, all admissible
weights at level
$k$ are generated by $\y.[\la^0+(k-k^0)\Lambda_0]$ with $\y$ and $\la^0$
spanning respectively the  sets  $\Y_{[u]}$ and 
$P^{k^0}/\W_+$.  Hence, all coset fields with weights having a nonvanishing
finite fractional part are identified with cosets with zero finite
fractional part.  This completes the argument of \msw. Moreover, it shows
that we can restrict to coset field representatives that have
$\y=id$, a point that was conjectured in \msw.  As a result, we can
always take as labels for coset fields, triplets of the form $\{\eta,\La;
\La'\, \}$. That is, we can choose  $\y=id$, with $\eta\in P_+^{\ell}, \la^0\in
P_+^{k^0}$ and $\la'^0\in P_+^{k^0+\ell u}$, and keep as the only residual field
identifications those generated by outer automorphisms:
\eqn\fiannm{\{\eta,\La;\La'\, \}\sim \{A\eta,A\La;A\La'\, \}\ .}
The entire analysis can therefore be completed in the $\y=id$ sector. This is
the  canonical description of the primary fields in these diagonal coset
theories.

It might appear somewhat more natural to reformulate the above result in 
terms of
finite weights ${\cal P} \la^0$ and ${\cal P} {\la'}^0$ since these are
integrable (i.e., elements of ${\bar P}_+^{k^0}$) in the
${\tilde y}=id$ sector. The residual field
identification then reads 
\eqn\fiannm{\{ {\cal P}{\eta},{\cal P} {\la}^0;{\cal P} {\la'}^0\, \}\sim 
\{ {\cal
P}A^u{ \eta},{\cal P}(A{\la}^0); {\cal P}(A{ \la'}^0)\, \}\ .}
The power $u$ of $A$ is simply justified by considering the branching
conditions.

\vfill\break

\newsec{Conclusion}

We have presented a simple, detailed description of the
Kac-Wakimoto principal admissible representations, together with two applications
in conformal field theory: 1 -- an elegant
transcription of the Malikov-Feigen-Fuchs null vectors and 2 -- the complete
analysis of the problem of field identifications in nonunitary diagonal cosets.

Coset models in which there are field
identifications other than those generated by outer automorphisms were 
thought to be extremely rare. Examples are the trivial coset models 
associated with 
conformal embeddings, as well as the maverick cosets found in \DJ\ (with a
further example noticed in \ref\fsc{J. Fuchs, B. Schellekens and C.
Schweigert, {\sl Nucl. Phys.} {\bf B461} (1996) 371}).  As we have shown,
nonunitary coset models are also special in that respect; they are mavericks. 
Of course, maverick field identifications seem to be a generic feature of
nonunitary cosets, whereas they are uncommon in unitary ones. 

The results of \mw\ \msw\ and of the present paper  constitute  the
coset counterpart to those obtained \ref\fkw{E. Frenkel, V. Kac and M. Wakimoto,
{\sl Comm. Math. Phys.} {\bf 147} (1992) 295} from the hamiltonian-reduction
point of view. In the few examples where the correspondence between the two
methods is known, they yield equivalent results. It should be stressed,
however, that the relation between hamiltonian reduction and coset theories
is not worked out in general (even though both constructions may be realised
as gauged WZW models \ref\gwzw{see, e.g.:\ L. Feher et al, {\it On the general
structure of Hamiltonian reductions of the WZNW theory},
hep-th/91112068; K. Gawedzki and A. Kupiainen, {\sl Phys. Lett.}
{\bf B215} (1988) 119}). For instance, no coset description is known for some
$W$-algebra models constructed by hamiltonian reduction, and the existence of
such coset descriptions has not yet been settled. This is
one of our motivations for pursuing the development of coset models.  Also, we
think that the coset description has advantages over the method of hamiltonian
reduction in the analysis of some properties, fusion rules being one example. 

We hope to return to this question of fusion rules for
(the conjectured) fractional-level WZW models. Our aim is to unravel the
Verlinde formulation of the null-vector expression for fusion rules and
to confront the different results\foot{Notice also that the results derived in
\bef\ using cohomology  and  in \ref\dlm{C. Dong, H. Li
and G. Mason, {\sl Comm. Math. Phys.} {\bf 184} (1997) 65} using a  
vertex-operator algebra agree with those derived by the Verlinde formula when
the latter yields positive fusion coefficients. So they differ from those
found in  \ay.}. One tactic would be to study their application to coset models,
where the final structure of fusion rules ought to be unambiguous.

\vskip 1truecm
\noindent{\it Acknowledgements}

For helpful conversations, M.W. thanks Wolfgang Eholzer, Matthias Gaberdiel, 
Terry Gannon, Adrian
Kent and G\'erard Watts; and acknowledges the
hospitality of D.A.M.T.P., Cambridge, where some of this work was done.

\bigskip
\noindent{\bf Appendix: Principal Admissible Weights for $C_2$ and $G_2$}
\bigskip

In this appendix, we describe the set $Y_{[u]}$ for the rank-two simple
algebras other than $A_2$.  Let us start with $C_2$. The list of all elements 
of $Y_{[4]}$ can be read off Figure 4 as those points in the fundamental domain
limited by the dark lines.  The different dashed lines indicate the reduction
appropriate to smaller values of $u$. The two types of octagons reflect the Weyl
group relations
$(r_0r_1)^4= (r_1r_2)^4=1$ while the diamond codes the simpler relation
$(r_0r_2)^2=1$.  The whole set of elements of ${\tilde Y}_{[u]}$ is obtained by 
the adjunction of the action of $\tilde W_+$ to the elements of ${Y}_{[u]}$.
For $C_2$, $\tilde W_+$ is isomorphic to $\Z_2$ and its nontrivial element
$A$ has action $A\al_0=\al_2 $, $A\al_2 = \al_0$.  Here is their
decomposition in the form
$t_\beta
\bar y$ for the case $u=3$:
\eqn\aaas{ \eqalignD{
&id = t_{(0,0)}\,id\qquad  &A= t_{(1,0)}\,r_2r_1r_2\cr
&r_0 = t_{(2,0)}\,r_1r_2r_1\qquad  &Ar_0= t_{(2,-1)}\,r_1r_2\cr
&r_1r_0 = t_{(-2,2)}\,r_2r_1\qquad  &Ar_1r_0= t_{(-2,1)}\,r_2\cr
&r_0r_1r_0 = t_{(0,2)}\,r_2r_1r_2\qquad & Ar_0r_1r_0= t_{(0,-1)}\,id\cr
&r_2r_1r_0 = t_{(2,-2)}\,r_1\qquad  &Ar_2r_1r_0= t_{(2,1)}\,r_2r_1r_2r_1\cr
&r_2r_0r_1r_0 = t_{(4,-2)}\,r_1r_2\qquad  &Ar_2r_0r_1r_0= t_{(4,-1)}\,r_1r_2r_1\cr
&r_1r_2r_1r_0= t_{(-2,0)}\,id\qquad  &Ar_1r_2r_1r_0= t_{(-2,3)}\,r_2r_1r_2\cr
&r_1r_2r_0r_1r_0 = t_{(-4,2)}\,r_2\qquad  &Ar_1r_2r_0r_1r_0 =
t_{(-4,3)}\,r_2r_1\cr &r_2r_1r_2r_0r_1r_0 = t_{(0,-2)}\,id\qquad 
&Ar_2r_1r_2r_0r_1r_0 = t_{(0,3)}\,r_2r_1r_2\cr}}
The values of $\beta$ span the set $-3<(\beta, \alpha)\leq 3$ with $\alpha\in
\{\al_1, \al_2, \al_1+\al_2, 2\al_1+\al_2\, \}$ and  $\beta\in
\bar P^\vee$, meaning that $\beta= (\beta_1,\beta_2)$ with $\beta_1$ even.  
The set ${\tilde
Y}_{[2]}$ is obtained by taking the first four elements in each column.
\midinsert
\vskip.25cm
\epsfxsize=7cm
\centerline{\epsfbox{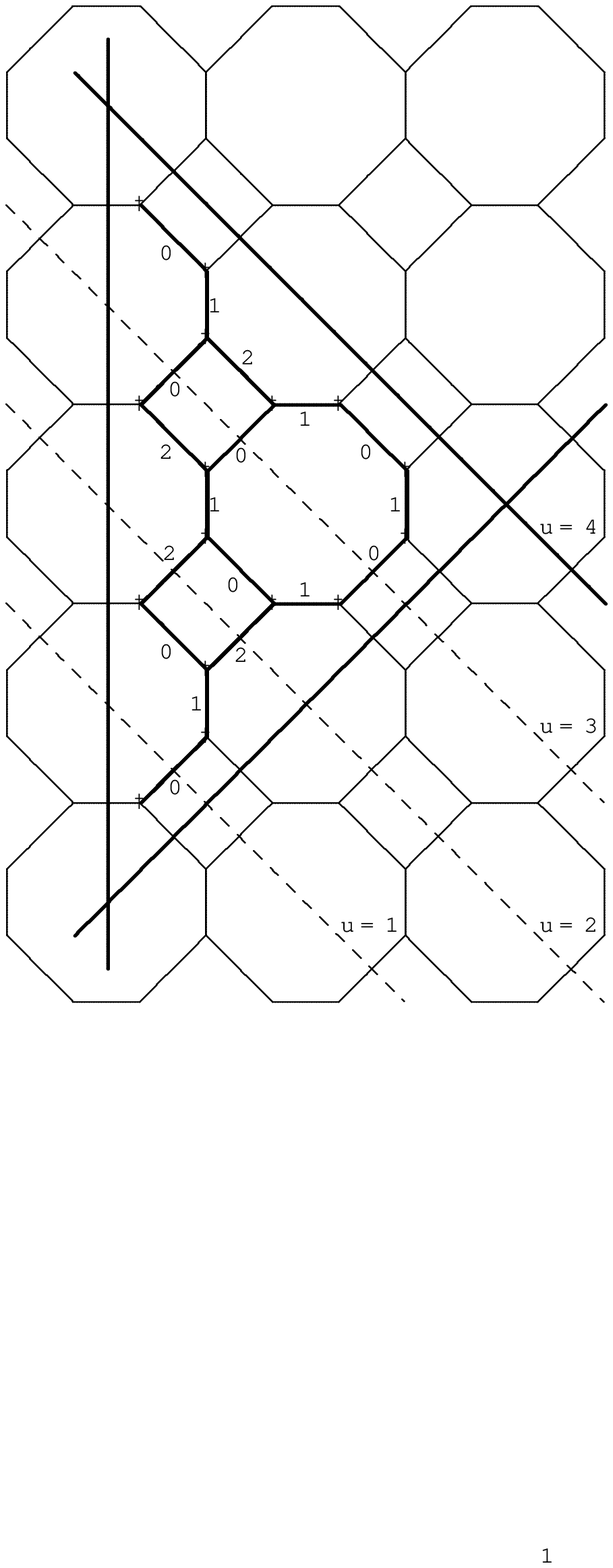}}
\vskip-4cm
\leftskip=1cm
\rightskip=1cm
\noindent
\baselineskip=12pt
{\it Figure 4}. The graph indicating the
elements of $Y_{[4]}$ for $C_2$. \bigskip\leftskip=0cm\rightskip=0cm
\baselineskip=15pt 
\endinsert 

Consider now $G_2$.  The diagrammatic description of the elements of $Y_{[u]}$ is
shown in Figure 5. Here the dodecagon, the hexagon and the
rectangle  translate the relations
$(r_1r_2)^6=(r_0r_1)^3= (r_0r_2)^2=id$, respectively. Every point within the
region bounded by the dark lines is associated to an element of  $Y_{[4]}$ and
again the dashed lines indicated the truncations resulting from decreasing
$u$.  The  decomposition of the elements of $Y_{[3]}$ in the form
$t_\beta
\bar y$ read
\eqn\aaas{ \eqalign{
&id = t_{(0,0)}id\cr
&r_0 = t_{(1,0)}r_1r_2r_1r_2r_1\cr
&r_1r_0 = t_{(-1,3)}r_2r_1r_2r_1\cr
&r_2r_1r_0 = t_{(2,-3)}r_1r_2r_1\cr
&r_1r_2r_1r_0 = t_{(-2,3)}r_2r_1\cr
&r_2r_1r_2r_1r_0 = t_{(1,-3)}id\cr
&r_0r_1r_2r_1r_0= t_{(0,3)}r_2r_1r_2r_1r_2\cr
&r_1r_2r_1r_2r_1r_0 = t_{(-1,0)}id\cr
&r_2r_0r_1r_2r_1r_0 = t_{(3,-3)}r_1r_2r_1r_2\cr}}
The values of $\beta$ span the set $-3<(\beta, \alpha)\leq 3$ with $\alpha\in
\{\al_1, \al_2, \al_1+\al_2, \al_1+2\al_2, \al_1+3\al_2, 2\al_1+3\al_2\, \}$ and  
$\beta= (\beta_1,\beta_2)$ with $\beta_2\in 3\Z$.  If we restrict to the first
four element above, we get ${\tilde Y}_{[2]}=Y_{[2]}$.  
 
Notice that for these two examples, there is no simple characterisation of
the elements of $Y_{[u]}$ in terms of the length of $y$.
\midinsert
\vskip.25cm
\epsfxsize=7cm
\centerline{\epsfbox{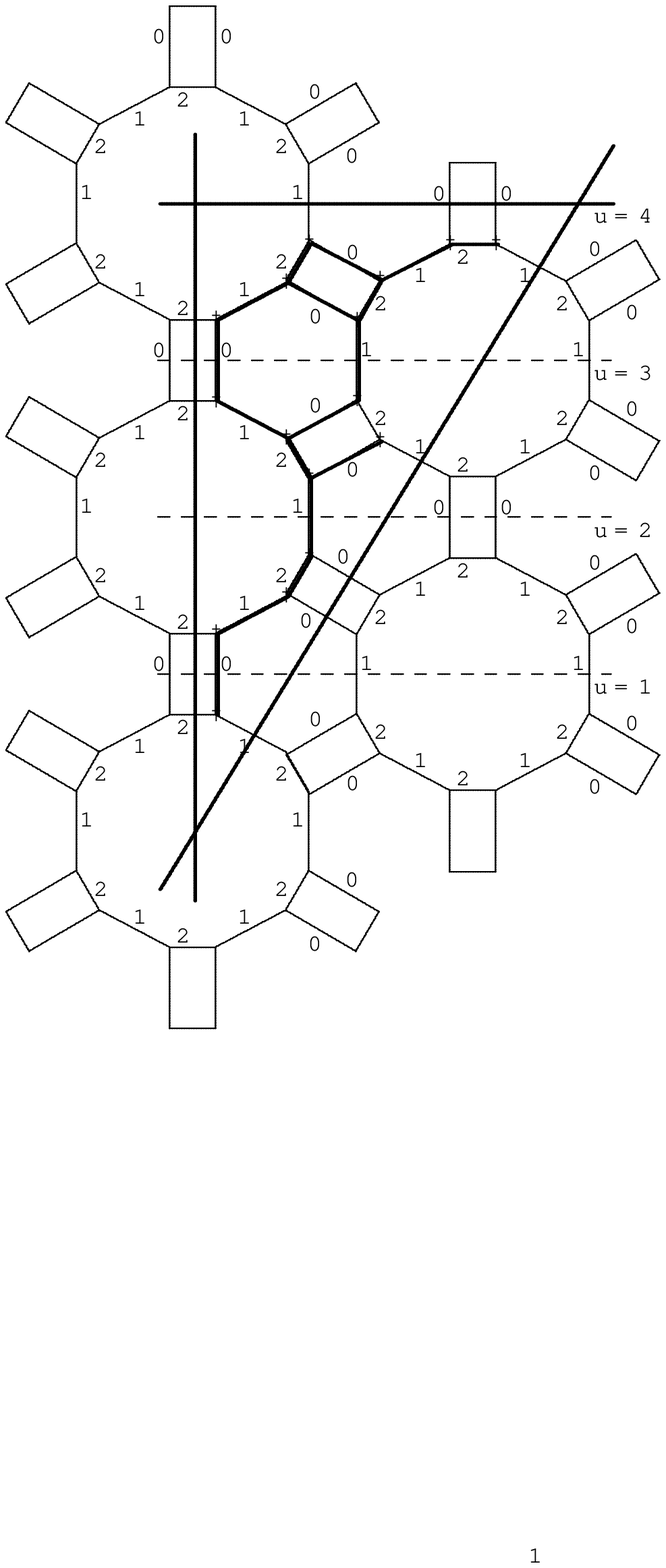}}
\vskip-4cm
\leftskip=1cm
\rightskip=1cm
\noindent
\baselineskip=12pt
{\it Figure 5}. The graph indicating the
elements of $Y_{[4]}$ for $G_2$. \bigskip\leftskip=0cm\rightskip=0cm
\baselineskip=15pt 
\endinsert

\listrefs

\bye